\documentclass[onecolumn,floatfix,11pt,nofootinbib]{revtex4}
\usepackage{amssymb,amsmath}
\usepackage{graphicx,float}
\usepackage{indentfirst}
\newcommand{\be}{\begin{equation}}
\newcommand{\ee}{\end{equation}}

\newcommand{\ba}{\begin{eqnarray}}
\newcommand{\ea}{\end{eqnarray}}

\begin{document}

\title{Quasi-homologous spherically symmetric branes and their symmetry breaking}

\author{M. C. B. Abdalla}
\email{mabdalla@ift.unesp.br}
\affiliation{Instituto de F\'isica Te\'orica, UNESP - Universidade Estadual Paulista,
Rua Dr. Bento Teobaldo Ferraz 271, Bloco II, Barra-Funda,
Caixa Postal 70532-2, 01156-970, S\~ao Paulo, SP, Brazil.}

\author{P. F. Carlesso}
\email{pablofisico@ift.unesp.br}
\affiliation{Instituto de F\'isica Te\'orica, UNESP - Universidade Estadual Paulista,
Rua Dr. Bento Teobaldo Ferraz 271, Bloco II, Barra-Funda,
Caixa Postal 70532-2, 01156-970, S\~ao Paulo, SP, Brazil.}

\author{J. M. Hoff da Silva}
\email{hoff@feg.unesp.br; hoff@ift.unesp.br}
\affiliation{Departamento de F\1sica e Qu\1mica, UNESP - Universidade
Estadual Paulista, Av. Dr. Ariberto Pereira da Cunha, 333,
Guaratinguet\'a, SP, Brazil.}


\begin{abstract}
We revisit the dynamical system based approach of spherically symmetric vacuum braneworlds, pointing out and studying the existence of a transcritical bifurcation as the dark pressure parameter changes its sign, we analyze some consequences of not discard the brane cosmological constant. For instance, it is noteworthy that the existence of an isothermal state equation between the dark fluid parameters cannot be obtained via the requirement of a quasi-homologous symmetry of the vacuum.
\end{abstract}

\maketitle


\section{Introduction}

Since 1999, braneworld models have attracted much attention of the scientific community. In fact, the view of our universe as a four-dimensional brane embedded into a five dimensional warped bulk \cite{RS} leads to new insights concerning well posed problems from particle physics to cosmology \cite{cosmo}. Several aspects of the braneworld cosmology were scrutinized, such as the gravitational collapse \cite{BH}, the cosmological dynamics \cite{DYN}, inflation \cite{INF} and cosmological perturbations \cite{PERT} (for a broader review, see \cite{MAA}). A quite interesting approach, opening several new perspectives was developed in Refs. \cite{jap,ali}. In fact, these works lead to an effective braneworld gravitational equation encoding local as well as non-local (purely geometrical) corrections. The non-local corrections are given by the explicit appearance of a specific bulk Weyl tensor projection.

In order to study the non-local contribution of the resulting gravitational equation, it is usual to parameterise the Weyl tensor by a type of (dark) cosmological fluid, respecting all the necessary constraints of the projection procedure. An analysis of spherically symmetric branes in the vacuum state,  specially devoted to some interesting brane symmetries investigation, was performed in \cite{HARMAK}. In particular, in this last paper, the notion of quasi-homologous branes, i. e., branes invariant with respect to the group of quasi-homologous transformations, is presented. In Refs. \cite{1,2,3}, a qualitative analysis, based upon the dynamical system defined by the projected equations, was carried out.

In this paper we aim to investigate first some qualitative aspects of quasi-homologous spherically symmetric vacuum braneworlds without an effective cosmological constant. To some extent it was previously developed \cite{HARMAK}, nevertheless we shall revisit this program, calling attention to the existence of a transcritical bifurcation as the dark pressure parameter is swept. This analysis is performed in Section III, being Section II devoted to some review and to present the first steps towards the structure equations. Then we move forward, rewriting the vacuum structure equations, this time with a brane cosmological constant. It is shown in Section IV that, in this last case, the effective cosmological constant, although quite small, may lead to important consequences in the braneworld picture. In particular, it is not possible to ensure an isothermal equation of state for the dark fluid by requiring a quasi-homologous invariant vacuum. Two aspects, in this research line, shall be mentioned. Firstly, an argument  in favor of keeping the cosmological constant term in the analysis. The usual approach is to discard the contributions coming from $\Lambda$. In fact, keeping in mind the minuteness of the brane (4D) cosmological constant, its effects can be safely neglected in finite sized gravitational systems as stars, galaxies, etc. However, since we are dealing with the vacuum brane itself, giving up of the brane cosmological constant terms seems to be an oversimplification. We note, in advance, the appearance of several terms involving a product of $\Lambda$ with the radial coordinate in the formulae (\ref{10})-(\ref{13}) below, making expected some effect in regions where $r$ is very large. Secondly, the fact that an isothermal equation of state for a cosmological fluid is consequence of a quasi-homologous symmetry (for usual systems) was reported long ago \cite{COLL} and translated to the braneworld language more recently \cite{HARMAK}. Hence, this type of symmetry breaking due to the presence of the cosmological constant is an important improvement in the braneworld scenario study, as it may (potentially) lead to new insights concerning the right relation between the dark pressure and energy.

\section{The structure equations}

It is well known that the projection of the 5D gravitational field equations on the brane leads to important corrections classified in two classes: the local corrections, encoded in the quadratic brane stress tensor terms, and the non-local (purely geometric) ones, encrypted in a specific projection of the bulk Weyl tensor. More precisely, being the five dimensional Einstein equations given by
\be G_{IJ}=\kappa_{5}^{2}T_{IJ}, \label{1}\ee where $G_{IJ}$ is the Einstein tensor, $\kappa_{5}^{2}=8\pi G_{5}$, and $T_{IJ}=\Lambda_{5}g_{IJ}+\delta(y)[-\lambda_{b} g_{IJ}+\tau_{IJ}]$. It is possible to show that the Gauss-Codazzi procedure leads to the following gravitational equations on an $y$ constant hypersurface \cite{jap}  \be G_{\mu\nu}=-\Lambda g_{\mu\nu}+\kappa_{4}^{2}\tau_{\mu\nu}+\kappa_{5}^{4}S_{\mu\nu}-E_{\mu\nu}. \label{2}\ee The notation is now evident: capital Latin indices go from $0..4$, while Greek indices vary in the range $0..3$. In the above expression, $S_{\mu\nu}$ is the aforementioned tensor encompassing quadratic brane stress tensor terms, as our analysis is in the brane vacuum this tensor vanishes. Before exploring the Weyl tensor $E_{\mu\nu}$ however, it is important to remark that after the projection procedure the constants $\Lambda$ and $\kappa_{5}$ are no longer fundamental. Instead, they are given by $\Lambda=\frac{\kappa_{5}^{2}}{2}(\Lambda_{5}+\kappa_{5}^{2}\lambda_{b}^{2}/6)$ and $\kappa_{4}^{2}=\kappa_{5}^{4}\lambda_{b}/6$, where $\lambda_{b}$ is the brane tension. In the vacuum, the Eq. (\ref{2}) reduces to \be R_{\mu\nu}=-E_{\mu\nu}+\Lambda g_{\mu\nu}\label{3}. \ee

As the standard approach to the Weyl tensor, its is parameterized in the following way \cite{RM} \be E_{\mu\nu}= \Bigg(\frac{\kappa_{5}}{\kappa_{4}}\Bigg)^{4}\Bigg[U\Big(u_{\mu}u_{\nu}+\frac{h_{\mu\nu}}{3}\Big)+2Q_{(\mu}u_{\nu)}+P_{\mu\nu}\Bigg]. \label{4} \ee This specific form for $E_{\mu\nu}$ is motived by the symmetries that it must obey. In Eq. (\ref{4}) $u^{\mu}$ is a given four velocity field and $h_{\mu\nu}=g_{\mu\nu}+u_{\mu}u_{\nu}$. It is possible to show \cite{HARMAK} that for a static spherically symmetric vacuum on the brane, $Q_{\mu}=0$ and $P_{\mu\nu}=P(r)(r_{\mu}r_{\nu}-h_{\mu\nu}/3)$, where $r_{\mu}$ is the unit radial vector and $r$ the radial distance. Besides, in an inertial frame on the brane one has $u^{\mu}=(1,\vec{0})$ and  $h_{\mu\nu}=diag(0,1,1,1)$. In the current jargon the decomposition (\ref{4}) is the so-called Weyl fluid, $U=U(r)$ is the dark radiation, and $P(r)$ is said as the dark pressure. This function will play an important role in the qualitative analysis to be presented in the next section.

After these introductory review we shall reobtain the structure equations for this case without discarding the brane cosmological constant. Using the usual static spherically symmetric line element on the brane \be ds^{2}=-e^{\nu(r)}dt^{2}+e^{\lambda(r)}dr^{2}+r^{2}(d\theta ^{2}+sin^{2}\theta d\phi ^{2}),\label{5} \ee we have

\be -e^{-\lambda}\Bigg(\frac{1}{r^{2}}-\frac{\lambda'}{r}\Bigg)+\frac{1}{r^{2}}=3\alpha U+\Lambda,\label{6} \ee

\be e^{-\lambda}\Bigg(\frac{\nu'}{r}+\frac{1}{r^{2}}\Bigg)-\frac{1}{r^{2}}=\alpha (U+2P)-\Lambda,\label{7} \ee

\be \frac{e^{-\lambda}}{2}\Bigg(\nu''+\frac{\nu'^{2}}{2}+\frac{\nu'-\lambda'}{r}+\frac{\nu'\lambda'}{2}\Bigg)=\alpha(U-P)-\Lambda,\label{8} \ee

\be \nu'=-\frac{U'+2P'}{2U+P}-\frac{6P}{r(2U+P)}, \label{9}\ee where $\alpha=\frac{16\pi G\kappa_{4}^{4}}{\kappa_{5}^{4}\lambda_{b}}$ and a prime denotes derivation with respect to $r$. From Eq. (\ref{6}) one can see that \be e^{-\lambda}=1-\frac{C}{r}-\frac{Q}{r}-\frac{\Lambda r^{2}}{3},\label{10} \ee being $Q=3\alpha \int Ur^{2}dr$. Substituting Eq. (\ref{9}) in (\ref{7}) one has \be U'=-2P'-\frac{6P}{r}+\frac{1}{r}(1-e^{\lambda})(2U+P)-(2U+P)re^{\lambda}[\alpha(U+2P)-\Lambda] \label{11} \ee and, with the aid of Eq. (\ref{10}), one arrives at the following expression \be \frac{dU}{dr}=-2\frac{dP}{dr}-6\frac{P}{r}-\frac{(2U+P)\Big(C+Q+\alpha r^{3}(U+2P)-\frac{2}{3}\Lambda r^{3}\Big)}{r^{2}\Big(1-\frac{C}{r}-\frac{Q}{r}-\frac{\Lambda r^{2}}{3}\Big)}.\label{12} \ee This last equation, together with $\frac{dQ}{dr}=3\alpha r^{2}U$, will be extensively studied in the following Sections.

\section{Quasi-homologous branes: Dynamical System analysis}

In order to set up a dynamical system analysis of the vacuum equations let us define \be q=\frac{C}{r}+\frac{Q}{r}+\frac{\Lambda r^{2}}{3}, \nonumber \ee  \be \mu=3\alpha r^{2}U+\Lambda r^{2},\nonumber \ee \be  p=3\alpha r^{2}P-2r^{2}\Lambda, \label{13}\ee together with the change of coordinate $\theta=ln(r)$. It can be readily verified that \be \frac{dq}{d\theta}=\mu-q,\label{14} \ee while a bit of algebra leads to \be \frac{d\mu}{d\theta}=2\mu-2p-2\frac{dp}{d\theta}-12e^{2\theta}\Lambda - \frac{(2\mu+p)[(\mu+2p)/3+q]}{1-q}.\label{15} \ee We note that, due to the presence of the cosmological constant term, the equations (\ref{14}) and (\ref{15}) are slightly different from the previously ones obtained in the literature.

After all the comprehensive discussion presented before \cite{HARMAK,3}, we would like to revisit the qualitative analysis of Eqs. (\ref{14}) and (\ref{15}), discussing another way to reach a quasi-homologous system, as well as interpreting the behavior of the fixed points. In order to accomplish that, we shall disregard the cosmological constant term along the present Section. As mentioned in the Introduction, when dealing with the vacuum of the brane itself, it seems to be an oversimplification. Even so, one may acquire physical insight by performing such a simplified analysis. Note that in order to investigate the dynamical system associated to the full equations (\ref{14}) and (\ref{15}) it would be necessary to prove the very existence of an absorbing set, which implies there exists a pullback attractor \cite{RASM}. It will not be done in this paper and we postpone to the next Section the importance of the cosmological constant for such a system.

Without the brane cosmological constant term, Eqs. (\ref{14}) and (\ref{15}) provide an useful starting point for the dynamical system analysis. Note, however, that the $dp/d\theta$ term is an obstruction to the dynamical system program (it is well known that dynamical systems in three or more dimensions may behave exotic. For instance, it was demonstrated in \cite{DIN} that the generic behavior of three dimensional trajectories approachs some strange attractor, although all the possible behaviors were not mapped yet). Generally, the approach used to overcome this problem is to implement {\it ab initio} an equation of state between $U$ and $P$ ($\mu$ and $p$). This procedure is closely related to the existence of a quasi-homologous symmetry on the brane. In fact, in Ref. \cite{HARMAK} it was stated that a theorem asserting that the vacuum brane equations (in the case we are dealing with, i. e., spherically symmetric and static) are invariant with respect to the group of quasi-homologous transformations if and only if $P=\gamma U$, where $\gamma$ is a constant. We shall comment more on that theorem in the next Section. Our procedure now is, instead of implementing the quasi-homologous symmetry as an input, to require a constant $p$, since in this way we immediately get an autonomous dynamical system out of (\ref{14}) and (\ref{15}). Obviously, it is also an assumption, and at first sight it is not related to the usual one ($P=\gamma U$) which engenders the quasi-homologous symmetry. It is possible to see, however, that the underlying dynamical system leads to the case $P=\gamma U$.

In order to accomplish that, we rewrite the structure equations with all the simplifications taken into account. Eq. (\ref{14}) remains unchanged, where now $q=\frac{C+Q}{r}$ and $\mu=3\alpha r^{2}U$, and \be \frac{d\mu}{d\theta}=2\mu-2p - \frac{(2\mu+p)[(\mu+2p)/3+q]}{1-q},\label{16} \ee being $p=3\alpha r^{2}P$. Notice that the assumption that $p$ is constant means $P\sim 1/r^{2}$. Let us write $p=3\alpha\beta$ with $\beta$ constant $(P=\beta/r^{2})$. Eq. (\ref{14}) trivially gives a strong constraint which must be respected by every single fixed point in the $(q,\mu)-$plane, namely $\mu_{*}=q_{*}$ where $*$ labels a fixed point. The fixed points are completely obtained from Eq. (\ref{16}). Hence \be \mu_{*}=\frac{3}{14}\Big[1-\alpha\beta \pm \sqrt{1-3\alpha\beta (9\alpha\beta+10)}\Big].\label{17} \ee We note from (\ref{17}) that \be \alpha\beta(9\alpha\beta+10) \leq 1/3,\label{18} \ee in order to have a real fixed point. This constraint together with the fact that $\mu_{*}\neq 1$ (see Eq. (\ref{16})), will play an important role in the subsequent analysis.

Notice that from $\mu_{*}=q_{*}$ we have \be 3\alpha r^{2}U=\frac{C+Q}{r}.\label{19} \ee Nevertheless, by the definition of $Q$, we have \be 3\alpha r^{2}U=\frac{dQ}{dr},\label{20} \ee in such a way that \be r\frac{dQ}{dr}=C+Q.\label{21}\ee Taking the derivative on both sides of Eq. (\ref{21}) with respect to $r$, we get \be \frac{d^{2}Q}{dr^{2}}=0,\nonumber \ee resulting in \be \frac{dQ}{dr}=3\alpha r^{2}U=\delta,\label{22} \ee where $\delta$ is a constant. As $\mu_{*}=q_{*}$ for $P=\frac{\beta}{r^{2}}$ ($p$ constant) it is readly verified that \be P=\frac{3\alpha\beta}{\delta}U,\label{23} \ee which means the existence of the quasi-holonomic symmetry \cite{HARMAK}.

Now, following the standard classification of critical points \cite{halekoc}, it is possible to find that the eigenvalues of the Jacobian matrix are given by \be \lambda_{\pm}=\frac{1}{\mu_{*}-1}\Bigg[\frac{19\mu_{*}}{6}+\frac{5\alpha\beta}{2}-\frac{3}{2}\pm \sqrt{193\mu_{*}^{2}+570\mu_{*}\alpha\beta-6\mu_{*}+441\alpha^{2}\beta^{2}+18\alpha\beta+9} \Bigg].\label{24} \ee Hence it is necessary to look at the real part of $\lambda$ for all the cases, namely ($\mu_{*}^{+},\lambda_{+}$), ($\mu_{*}^{+},\lambda_{-}$), and ($\mu_{*}^{-},\lambda_{+}$), ($\mu_{*}^{-},\lambda_{-}$). The classification is quite characteristic \cite{halekoc}: in a given fixed point, if $Re(\lambda_{\pm})>0$, then the fixed point is a repellor; if  $Re(\lambda_{\pm})<0$, then the fixed point in question is an attractor. Finally, if $Re(\lambda_{+})$ and $Re(\lambda_{-})$ have opposite signs, then the fixed point is called a saddle point.  As mentioned, we shall make explicit the existence of a transcritical bifurcation in the dynamical system in question. In order to unveil this characteristic, we shall investigate the behavior (classification) of each critical point as the dark pressure parameter changes its sign. As we shall see, there is a complete change in the behavior of one of the critical points,
characterizing the aforementioned bifurcation.

\subsection{The $p=0$ case}

In order to get a physical insight on the possibilities raised in the scope of the dynamical systems behind the structure equations, let us investigate the simplest case with $p=0$ ($\beta=0$) in Eqs. (\ref{17}) and (\ref{24}). From (\ref{17}) it is simple to see that the critical points are $\mu_{*}^{+}=3/7$ and $\mu_{*}^{-}=0$. For the case $\mu_{*}^{+}=3/7$, the eigenvalues (\ref{24}) are given by \be \lambda_{+}=-\frac{1}{8}(-2+\sqrt{228})<0\label{25} \ee and \be \lambda_{-}=-\frac{1}{8}(-2-\sqrt{228})>0,\label{26}\ee therefore this point is a (less important) saddle point. The other critical point ($\mu_{*}^{-}=0$) has \be \lambda_{-}=2\lambda_{+}=2,\label{27} \ee hence it is a repellor. Note that from the constraint $\mu_{*}=q_{*}=p=0$ we need necessarily to have $C=0$. This fact is indeed technically sound, since in the absence of the dark fluid we should expect a complete vacuum on the brane, and $C=0$ (when comparing with the Schwarzschild case) shall be recognized as $2MG=0$. In this specific critical point, the gravitational equations give \be ds^{2}=-dt^{2}+dr^{2}+r^{2}(d\theta^{2}+sin^{2}\theta d\phi^{2}),\label{28}\ee with a constant absorbed in the $t$ coordinate. This ``spherical symmetry'', from the analysis of the dynamical system, is obviously unstable.

\subsection{The $p>0$ case}

The fisrt thing we shall note in this case is the constraint (\ref{18}). Since now $\beta>0$, the allowed range of values of $\alpha\beta$ is within the interval $\Big(0,\frac{-30+\sqrt{1008}}{54}\Big]$, stressing for the minuteness of possible values. Therefore, it is necessary to substitute the Eq. (\ref{17}) into (\ref{24}) analyzing all the possible cases. This approach is, however, a hard task. Hence it is useful to plot all the possible cases and, then, scrutinize the sign possible variations of the eigenvalues.

From the Figs. (1) and (2), it is possible to conclude that for the fixed point $\mu_{*}^{+}$ the eigenvalue $\lambda_{-}$ is always positive within the relevant range. Instead, $\lambda_{+}$ can be positive or negative. Therefore, $\mu_{*}^{+}$ is either saddle or repellor.

The situation is also simple for the $\mu_{*}^{-}$ fixed point. From Figs. (3) and (4) it can be readly verified that this point is always a repellor. Nevertheless, the situation can be modified if we consider the negative reduced dark pressure case.

\begin{figure}[H]
\begin{center}
\includegraphics[width=3.3in]{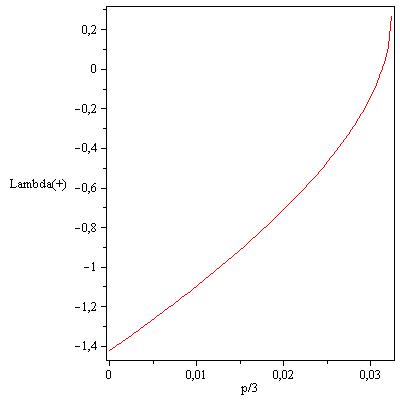}
\end{center}
\caption{\small Eigenvalue $\lambda_{+}$ for the fixed point $\mu_{*}^{+}$.}
\end{figure}

\begin{figure}[H]
\begin{center}
\includegraphics[width=3.3in]{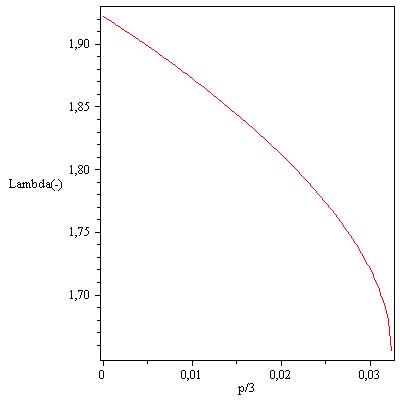}
\end{center}
\caption{\small Eigenvalue $\lambda_{-}$ for the fixed point $\mu_{*}^{+}$.}
\end{figure}

\begin{figure}[H]
\begin{center}
\includegraphics[width=3.3in]{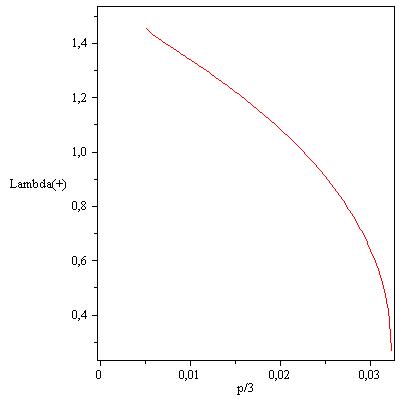}
\end{center}
\caption{\small Eigenvalue $\lambda_{+}$ for the fixed point $\mu_{*}^{-}$.}
\end{figure}

\begin{figure}[H]
\begin{center}
\includegraphics[width=3.3in]{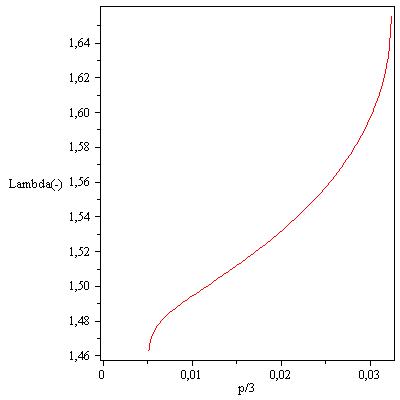}
\end{center}
\caption{\small Eigenvalue $\lambda_{-}$ for the fixed point $\mu_{*}^{-}$.}
\end{figure}

\subsection{The $p<0$ case}

The last possibility in the possible range of the reduced dark pressure is the negative case. In this case, the constraint (\ref{18}) reads \be \alpha|\beta|(10-9\alpha|\beta|)\geq-\frac{1}{3} ,\label{29}\ee thus we can see that the relevant range for this case is $\alpha |\beta| \in \Big(0,\frac{30+\sqrt{1008}}{54}\Big]$. Notice the peculiar behavior presented in Figs. (5) and (6). Inspite of the absence of similarity between the curves, there is an interesting pattern around the value 0.7 of the horizontal axis. The change of sign occurs at the same interval, in such a way that the critical point $\mu_{*}^{+}$ is always a saddle point. Another remarkable behavior is happening, this time with the point $\mu_{*}^{-}$. The Figs. (7) and (8), show that the eigenvalues at this last critical point are both positive until $p/3=1$, and therefore it is classified as a repellor. Notwithstanding, from this point on, both eigenvalues become negative, and the critical point $\mu_{*}^{-}$ starts to act as an attractor. Comparing with the $p>0$ case, in which this fixed point is always a repellor, now we have a drastic change of classification. This typical behavior is well known in dynamical system theory, and characterizes the so-called transcritical bifurcation.

\begin{figure}[H]
\begin{center}
\includegraphics[width=3.3in]{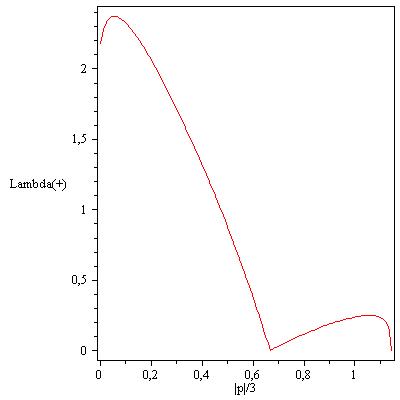}
\end{center}
\caption{\small Eigenvalue $\lambda_{+}$ for the fixed point $\mu_{*}^{+}$.}
\end{figure}

\begin{figure}[H]
\begin{center}
\includegraphics[width=3.3in]{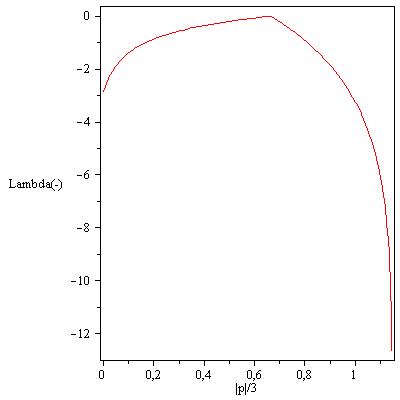}
\end{center}
\caption{\small Eigenvalue $\lambda_{-}$ for the fixed point $\mu_{*}^{+}$.}
\end{figure}

\begin{figure}[H]
\begin{center}
\includegraphics[width=3.3in]{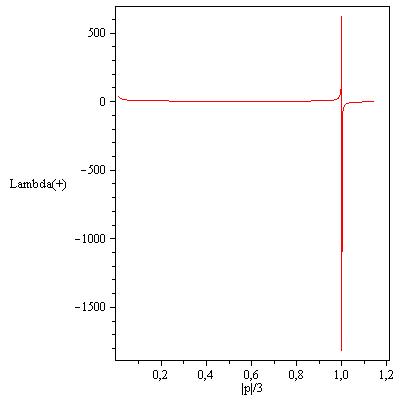}
\end{center}
\caption{\small Eigenvalue $\lambda_{+}$ for the fixed point $\mu_{*}^{-}$. In this plot, the horizontal axis has been shortened for the sake of exposition.}
\end{figure}

\begin{figure}[H]
\begin{center}
\includegraphics[width=3.3in]{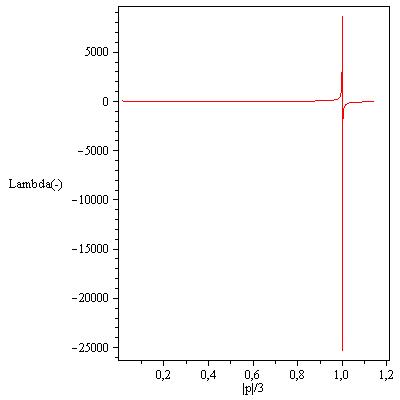}
\end{center}
\caption{\small Eigenvalue $\lambda_{-}$ for the fixed point $\mu_{*}^{-}$. In this plot, the horizontal axis has been shortened for the sake of exposition.}
\end{figure}

It may be interesting to summarize the situation: For the $p>0$ case, the fixed point $\mu_{*}^+$ is a repellor or a saddle point, while the fixed point $\mu_{*}^-$ is always a repellor. Instead, for the $p<0$ case, the fixed point $\mu_{*}^+$ is always a saddle point, while the fixed point $\mu_{*}^-$ is a repellor just for some values of $p$. For other values of $p$, $\mu_{*}^-$ change its behaviour, acting as a attractor. This change of behaviour is precisely what characterizes a transcritical bifurcation.  

Before going further it would be interesting to return to Equations (\ref{14}) and (\ref{15}) and see what possible influence -- if any -- the cosmological constant term may bring into light, when taken into account. In the numerical results below, since we do not want to assume any state equation to the dark sector a priori, we set $p=0$, simplifying the procedure. In all the figures the doted line stands for $q(r)$ while the dashed line means $\mu(r)$.

From the Figues (9) and (10) it is possible to see that for relatively small scales there is no difference between the case with and without the cosmological constant term. The situation, however, is completely different for huge scales, as shown in Figs. (11) and (12). As the cosmological constant term is small, its influence is only present at large scales, as expected. Note that the contrast evinced by the Figs. (11) and (12) is a cogent argument for the investigation of the braneworld gravitational equation in the presence of the cosmological constant term. 

\begin{figure}[H]
\begin{center}
\includegraphics[width=5.0in]{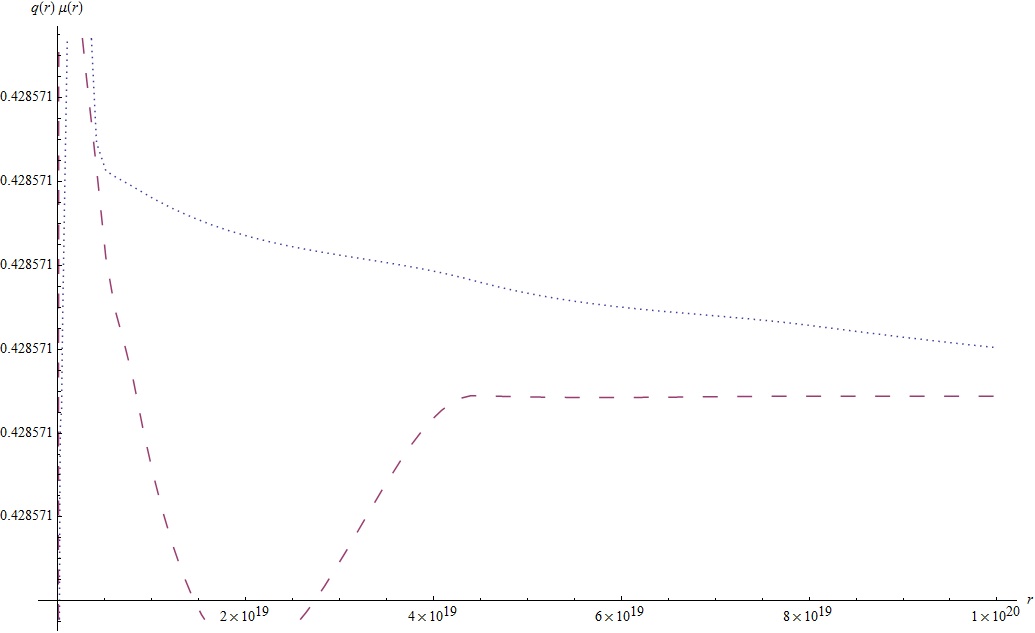}
\end{center}
\caption{\small The behaviour of $\mu$ and $q$ {\bf without} cosmological constant for relatively small scales. The doted line stands for $q(r)$ while the dashed line means $\mu(r)$. Note the small variance in the vertical axis.}
\end{figure}

\begin{figure}[H]
\begin{center}
\includegraphics[width=5.0in]{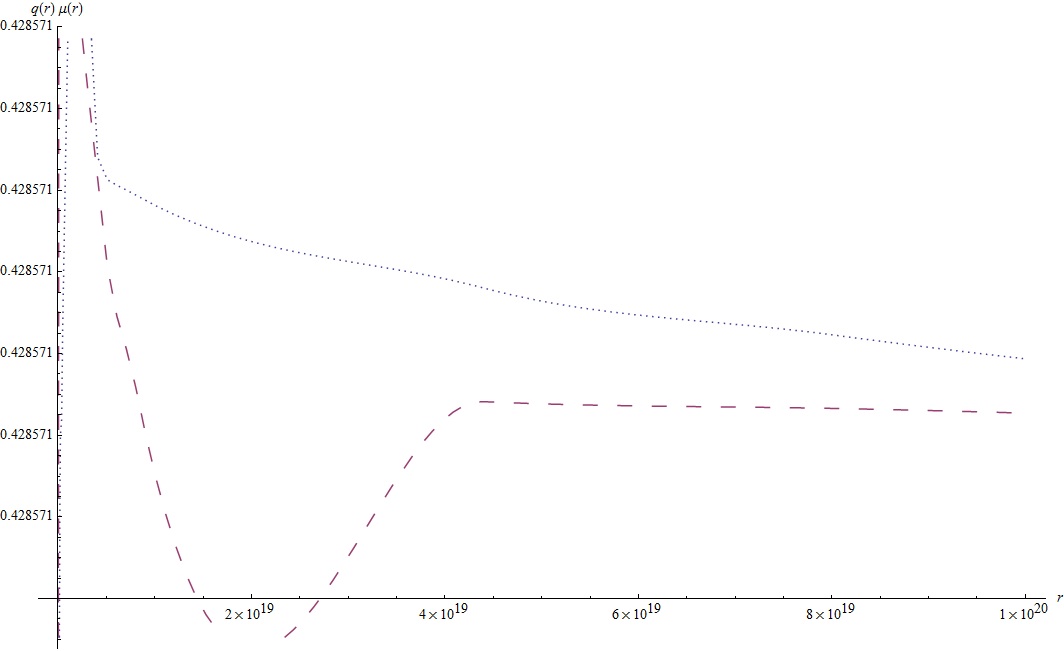}
\end{center}
\caption{\small The behaviour of $\mu$ and $q$ {\bf with} cosmological constant for relatively small scales. The doted line stands for $q(r)$ while the dashed line means $\mu(r)$. Note the small variance in the vertical axis.}
\end{figure}

\begin{figure}[H]
\begin{center}
\includegraphics[width=5.0in]{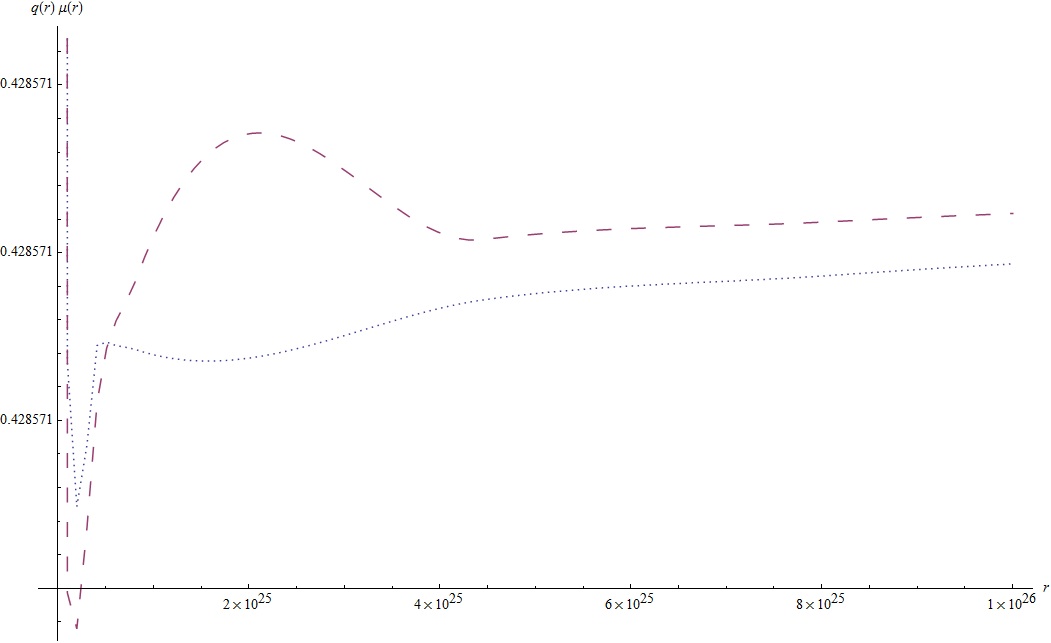}
\end{center}
\caption{\small The behaviour of $\mu$ and $q$ {\bf without} cosmological constant for large scales. The doted line stands for $q(r)$ while the dashed line means $\mu(r)$}
\end{figure}

\begin{figure}[H]
\begin{center}
\includegraphics[width=5.0in]{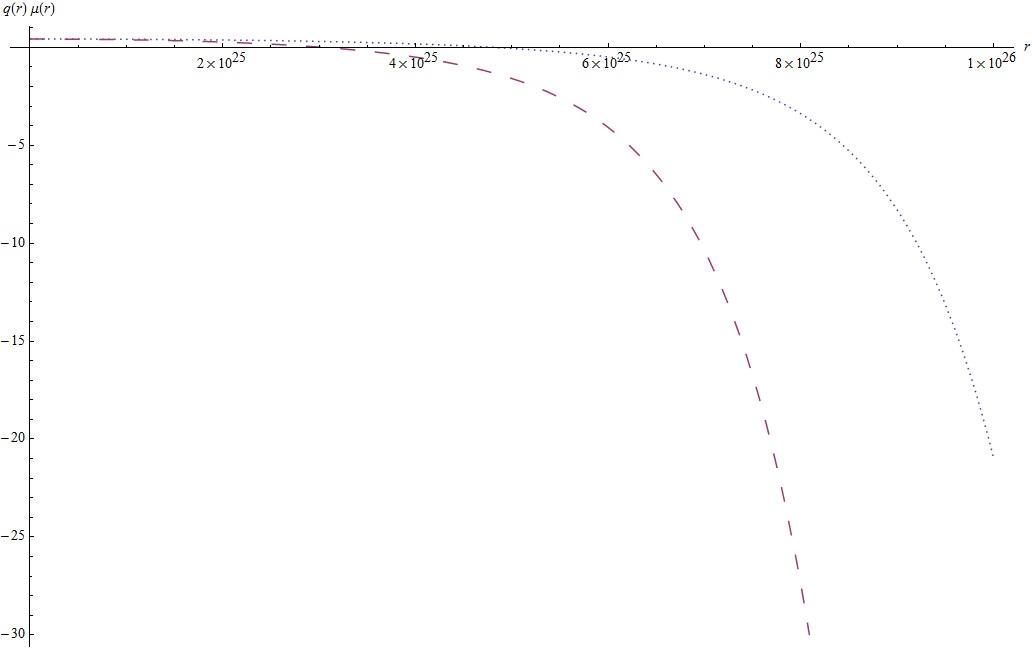}
\end{center}
\caption{\small The behaviour of $\mu$ and $q$ {\bf with} cosmological constant for large scales. The doted line stands for $q(r)$ while the dashed line means $\mu(r)$}
\end{figure}

\section{General system with brane cosmological constant: breaking usual quasi-homologous symmetry}

It was first noted by C. B. Collins \cite{COLL} that the arguments, somewhat {\it ad hoc}, leading to the equation of state of static (Newtonian and relativistic) stars may be obtained from the requirement that a (quasi) homologous family of solutions should exist for this problem. This concept was successfully extended to the (spherically symmetric and static) braneworld case in \cite{HARMAK}. In this last case it was shown that the infinitesimal operator generating the group of quasi-homologous transformations is given by \be {\bf H}=r\frac{\partial}{\partial r}-2U\frac{\partial}{\partial U}+(Q+C)\frac{\partial}{\partial Q}.\label{30}\ee Moreover, it is shown that the invariance with respect to this quasi-homologous transformations ensures the relation $P=\gamma U$, with $\gamma$ constant, and vice-versa \cite{HARMAK}.

We shall demonstrate here that the usual quasi-homologous symmetry is breaking due to the presence of the cosmological constant. Hence, in particular, the simple equation of state is not more sufficient to ensure the quasi-homologous symmetry. Before doing so, however, let us make a brief account on the aforementioned symmetry.

Generally speaking, it is said that a given system of (ordinary) differential equations \be \frac{du^{m}}{dt}=f^{m}(t,{\bf u}),\label{31} \ee being $m=1,..,n$ and ${\bf u}=(u^{1},..,u^{n})$ is invariant under quasi-homologous transformations if, and only if \be \frac{d\eta^{m}(u^{m})}{du^{m}}-\frac{d\pi(t)}{dt}={\bf \tilde{H}}(ln|f^{m}|),\label{32} \ee with no sum in the $m$ index. In the above expression, ${\bf \tilde{H}}=\pi(t, {\bf u})\frac{\partial}{\partial t}+\eta^{m}(t,{\bf u})\frac{\partial}{\partial u^{m}}$ is the infinitesimal generator whose action leaves the set (\ref{31}) invariant. The reason for the specific form of Eq. (\ref{32}) rests upon the application of Lie group technics in the investigation of differential equations systems and we refer the reader for more details, for instance, to the Ref. \cite{novo}. It is worth, however, giving a precise definition for quasi-homologous symmetry: in the above context it means symmetry under transformations as $t\rightarrow \tilde{t}(t)$ and $u^{m}\rightarrow \tilde{u}^{m}(u^{m})$. After all, the constraints imposed by the left hand side of Eq. (\ref{32}) on the allowable quasi-homologous transformations are such that the unique possibility is given by a simple rescaling \cite{COLL}. This fact will be important in the interpretation of quasi-homologous symmetry breaking.

Returning to our problem, let us assume that the dark pressure and the dark energy are related by the simple state equation $P(U)=\gamma U$, being $\gamma$ constant. Remember that, as mentioned, it is a sufficient condition to ensure quasi-homologous symmetry in the case without cosmological constant. The relevant vacuum gravitational equations are given by \be \frac{dQ}{dr}=3\alpha r^{2}U \label{33} \ee and \ba \frac{dU}{dr}&=&\left.\frac{-\gamma U}{(1+2\gamma)r^{2}\Bigg(1-\frac{C}{r}-\frac{Q}{r}-\frac{\Lambda r^{2}}{3}\Bigg)}\Bigg[(1+2\gamma^{-1})\Bigg(C+Q+\alpha r^{3}U(1+2\gamma)-\frac{2\Lambda r^{3}}{3}\Bigg)\right.\nonumber\\&+&\left.6r-6(C+Q)-\frac{6\Lambda r^{3}}{3} \Bigg].\right.\label{34} \ea Let us investigate the possibility of the usual quasi-homologous transformations generated by the, also usual, infinitesimal generator \be {\bf H}=\xi(r)\frac{\partial}{\partial r}+\eta^{1}(U)\frac{\partial}{\partial U}+\eta^{2}(Q)\frac{\partial}{\partial Q},\label{35}\ee deliberately taken as being the same of Refs. \cite{HARMAK,COLL} (it is in this sense we call it `usual'). Applying the generator (\ref{35}) in (\ref{33}) we have, from (\ref{32}), the result \be \frac{d\eta^{2}(Q)}{dQ}-\frac{d\xi(r)}{dr}=\frac{2\xi(r)}{r}+\frac{\eta^{1}(U)}{U},\label{36} \ee whose solution is given by \cite{HARMAK} \be \xi(r)=\frac{a}{r^{2}}+br,\label{37} \ee \be \eta^{1}(U)=(c-3b)U\label{38} \ee and \be \eta^{2}(Q)=cQ+d,\label{39} \ee where $a$ and $d$ are integration constants, whereas $c$ and $3b$ are separation constants.

The equation to be satisfied in the case of Eq. (\ref{34}) is more involved. The left hand side, given by $\frac{d\eta^{1}(U)}{dU}-\frac{d\xi(r)}{dr}$, is trivially obtained from Eqs. (\ref{37}--\ref{39}). It reads simply \be \frac{2a}{r^{3}}+c-4b. \label{40}\ee The complete Eq. (\ref{32}) in this case reads \ba  \frac{2a}{r^{3}}+c-4b&=&\left. \Big(\frac{a}{r^{2}}+br\Big)\Bigg[ \frac{(1+2\gamma^{-1})(1+2\gamma)3\alpha r^{2}U+6-4\Lambda r^{2}(2+\gamma^{-1})}{F_{\Lambda}}\right.\nonumber\\&-&\left.\frac{1+\Big(1-\frac{(C+Q)}{r}\Big)-\frac{4\Lambda r^{2}}{3}}{r\Big(1-\frac{(C+Q)}{r}\Big)-\frac{\Lambda r^{3}}{3}} \Bigg]+(c-3b)U\Bigg[\frac{1}{U}+\frac{\alpha r^{3}(1+2\gamma^{-1})(1+2\gamma)}{F_{\Lambda}}\Bigg]\right.\nonumber\\&+&\left. (cQ+d)\Bigg[\frac{2\gamma^{-1}-5}{F_{\Lambda}}+\frac{1}{r\Big(1-\frac{(C+Q)}{r}\Big)-\frac{\Lambda r^{3}}{3}}\Bigg]\right.,\label{41}\ea where $F_{\Lambda}=(1+2\gamma^{-1})[C+Q+\alpha r^{3}U(1+2\gamma)]\!+\!6r\!-\!6(C+Q)\!-\!\frac{4\Lambda r^{3}}{3}(2+\gamma^{-1})$. After a simple, but lengthy, algebra it is possible to rewrite Eq. (\ref{14}) as \ba &&\left. \frac{1}{1-\frac{C}{r}-\frac{Q}{r}-\frac{\Lambda r^{2}}{3}}\Bigg[\frac{3a}{r^{3}}\bigg(1-\frac{C}{r}-\frac{Q}{r}\bigg)-\bigg(\frac{2a}{r^{3}}+b\bigg)\Lambda r^{2}\!-\!\bigg(\frac{d}{r}+\frac{cQ}{r}\bigg)\!+\!\frac{a}{r^{3}}+b \Bigg]=\right.\nonumber\\&& \left. \frac{\bigg(\frac{3a}{r^{2}}+cr\bigg)(1+2\gamma^{-1})(1+2\gamma)\alpha r^{2}U+6\bigg(\frac{a}{r^{2}}+br\bigg)+(cQ+d)(2\gamma^{-1}-5)-4\Lambda r^{2}(2+\gamma^{-1})\bigg(\frac{a}{r^{2}}+br\bigg)}{(1+2\gamma^{-1})(1+2\gamma)\alpha r^{3}U+6r+(C+Q)(2\gamma^{-1}-5)-\frac{4\Lambda r^{3}}{3}(2+\gamma^{-1})}.\right.\nonumber\ea

It can be easily verified that, as expected, if $a=b=c=d=0$ the above constraint is satisfied. In fact, it corresponds to the identity transformation $({\bf H}=0)$. Another important remark is that if we take $\Lambda =0$, then the only possibility to satisfy the constraint is $a=0$, $d=C$ leading to $c=b=1$, which is exactly the solution found in Ref. \cite{HARMAK}, linking, in this way, the state equation with the generator (\ref{30}). Note, however, that the presence of the brane cosmological constant term prevents the existence of (integration and separation) constants that satisfy the equality, breaking the quasi-homologous symmetry. We shall interpret this fact as follows: as mentioned, the existence of a quasi-homologous symmetry means invariance under simple rescaling of the physical parameters. Nevertheless, the cosmological constant has a small fixed value. In the braneworld paradigm it depends on the brane tension and on the five-dimensional cosmological constant, both fixed parameters. Besides, this dependence is such that it enables a quite specific value for the brane cosmological constant. Therefore, from this perspective, the system of equations taking into account the cosmological constant term shall not be invariant under rescaling transformations.

\section{Concluding remarks}

We would like to emphasize in this concluding remarks the main accomplishes of this work. By investigating the subjacent dynamical system associated to the brane vacuum equations for the spherically symmetric and static case, we called the attention to the very existence of a transcritical bifurcation by the variation of the dark pressure parameter. It is important to stress that, in view of Section  III, the dark pressure parameter must be negative (the existence of the attractor point occurs only in this case). Besides, it was shown the possible values of $p$ by inspecting the necessary constraint (\ref{18}) to have a real fixed point in the configuration plane. It is important to call attention to the minuteness of $p$. Note also that in all the cases we have $\mu_{*}\neq 1$ for the allowed $p$ values, which is desirable for inner consistency of the dynamical system. Furthermore, we shown that the usual equation of state $P=\gamma U$ (or the existence of quasi-homologous symmetry) is also a consequence of the requirement $p$ constant. Generally speaking, it is well known that the effect of induced-gravity at early times is to restore the usual cosmological behavior of the universe. At late times, however, the results of general relativity are no longer recovered and the acceleration can be driven by extra-dimensional gravity effects \cite{timo}. What we see from our previous analysis is another facet of this behavior. The attractor point existence needs a negative dark pressure. After all, the situation is clear: the Weyl tensor precludes the existence of dark energy. If we parameterize it as a dark ``fluid'', then such a ``fluid'' must have negative pressure.  

Going further, it was demonstrated that the presence of the brane effective cosmological constant term ($\Lambda$) breaks the usual quasi-homologous symmetry. This symmetry breaking is due to the fact that $\Lambda$ is a fixed parameter. This is an important result, exemplifying the importance of not neglecting the cosmological constant. In other words, even being $\Lambda$ a quite small constant, its presence leads to new information about the vacuum brane system. We conclude emphasizing that the quasi-homologous symmetry could, in principle, be restored via an extension of the ${\bf H}$ generator, or by allowing a more elaborate relation between the dark pressure and dark energy. The program concerning the restauration of the quasi-homologous symmetry is potentialy interesting since it may, in principle, lead to new brane cosmology aspects based upon a novel relation between pressure and energy in the dark sector parameterized by the Weyl tensor. This is obviously a comprehensive program, being currently under investigation. 

\section*{Acknowledgments}

M. C. B. Abdalla thanks to Conselho Nacional de Desenvolvimento Cient\'{\i}fico e Tecnol\'ogico (CNPq) for financial support. J. M. Hoff da Silva thanks to Instituto de F\'isica Te\'orica and Niels Bohr Institute for hospitality during part of this work and to CNPq (482043/2011-3; 308623/2012-6). P. F. Carlesso thanks to CAPES-Brazil for financial suport.

\end{document}